\documentclass[aps,prl,twocolumn,superscriptaddress,preprintnumbers,amsmath,amssymb,floatfix,nofootinbib]{revtex4}
\usepackage{graphicx}
\usepackage{dcolumn}
\usepackage{bm}
\usepackage{url}
\usepackage{amsmath}
\usepackage{amssymb}
\usepackage{color}
\usepackage{subfig}
\usepackage{ulem}

\begin{document}

\title{$CP$ Asymmetries in Charm Decays into Neutral Kaons}
\author{Di Wang}
\affiliation{%
School of Nuclear Science and Technology,  Lanzhou University,
Lanzhou 730000,  People's Republic of China}
\author{Fu-Sheng Yu}\email{yufsh@lzu.edu.cn}
\affiliation{%
School of Nuclear Science and Technology,  Lanzhou University,
Lanzhou 730000,  People's Republic of China}
\author{Hsiang-nan Li}\email{hnli@phys.sinica.edu.tw}
\affiliation{%
Institute of Physics, Academia Sinica, Taipei, Taiwan 115, Republic of China}
 
\begin{abstract}

We find a new $CP$-violation effect in charm decays into neutral kaons, 
which results from the interference between two tree (Cabibbo-favored 
and doubly Cabibbo-suppressed) amplitudes with the mixing of final-state 
mesons. This effect, estimated to be of an order of $10^{-3}$, is much 
larger than the direct $CP$ asymmetries in these decays, but missed in 
the literature. It can be revealed by measuring the difference of the 
time-dependent $CP$ asymmetries in the $D^{+}\to \pi^{+}K_S^0$ and 
$D_{s}^{+}\to K^{+} K_S^0$ modes, which are accessible at the LHCb 
and Belle II experiments. If confirmed, the new effect has to be 
taken into account, as the above direct $CP$ asymmetries are used to
search for new physics. 

\end{abstract}

\maketitle

$CP$ violation plays an important role in interpreting the matter-antimatter asymmetry in the Universe and in searching for new physics beyond the Standard Model (SM). It has been well established in the kaon and $B$ meson systems, but not yet in the charm sector. Many theoretical and experimental efforts have been devoted to the study of $CP$ violation in 
the singly Cabibbo-suppressed (SCS) $D$ meson decays, with the interests in flavor-changing-neutral currents from penguin amplitudes. The most precise individual measurements up to now are obtained for the time-integrated $CP$ asymmetry by the LHCb Collaboration \cite{Aaij:2016cfh},
\begin{align}\label{data}
\Delta A_{CP} &\equiv A_{CP}(D^0\to K^+K^-)- A_{CP}(D^0\to \pi^+\pi^-)
\nonumber\\
&=(-1.0\pm0.8\pm0.3)\times10^{-3},
\end{align}
which is dominated by the direct $CP$ violation $\Delta a_{CP}^{\rm dir}$,
and for the asymmetry in effective decay widths through time-dependent rates \cite{Aaij:2017idz}, $A_{\Gamma}(D^{0}\to K^{+}K^{-})=(-0.30\pm0.32\pm0.10)\times10^{-3}$, which is sensitive to the indirect $CP$ violation in the $D^{0}$-$\overline D^{0}$ mixing. With the precision lower than $10^{-3}$, there is still no evidence of $CP$ violation in the charm system. 

$CP$ violation can also occur via the interference between the
Cabibbo-favored (CF) and doubly Cabibbo-suppressed (DCS) channels of
$D\to fK^0_{S}$, with $f$ being a final-state particle.
These decays, with large branching factions from the CF amplitudes, 
are more experimentally accessible. However, the $CP$ asymmetries, such as 
\begin{equation}\label{eq:expDp}
A_{CP}(D^+\rightarrow \pi^+K_S^0) = (-3.63\pm0.94\pm0.67)\times 10^{-3},
\end{equation}
with $3.2\sigma$ from zero observed by the Belle Collaboration \cite{Ko:2012pe}, are mainly attributed to the $K^0$-$\overline K^0$ mixing. It has been claimed \cite{Ko:2012pe,Lipkin:1999qz,DAmbrosio:2001mpr,Grossman:2011zk,Bianco:2003vb} that deviation from the kaon-mixing effect in a precise measurement of the above mode can be identified as the direct $CP$ violation. Because of its smallness in the SM, the direct $CP$ violation in these decays has been regarded as a promising observable for searching for new physics \cite{Bigi:1994aw,Xing:1995jg,Lipkin:1999qz,DAmbrosio:2001mpr}. 

In this Letter, we will point out a new $CP$-violation effect in charm decays into neutral kaons, which results from the interference between the CF and DCS amplitudes with the mixing of final-state mesons. This new effect, estimated to be of the order of $10^{-3}$,  turns out to be much larger than the direct $CP$ asymmetry, but has been, to our surprise, missed in the literature \cite{Lipkin:1999qz,Ko:2012pe,Grossman:2011zk,Bianco:2003vb}.  We propose to measure the difference of the $CP$ asymmetries in the decay chains $D^{+}\to \pi^{+}K(t)(\to \pi^+\pi^-)$ and $D_{s}^{+}\to K^{+}K(t)(\to \pi^+\pi^-)$, where $K(t)$ represents a time-evolved neutral kaon $K^0(t)$ or $\overline K^0(t)$ with $t$ being the time difference between the charm decays and the neutral kaon
decays in the kaon rest frame. It will be shown that the contributions
to the above difference from the pure kaon mixing cancel, and the 
new effect can be clearly revealed. Only when this new effect has been 
well determined, can the direct $CP$ asymmetries in charm decays into neutral kaons be extracted correctly and used to search for new physics.

A $K_S^0$ state is reconstructed via its decay into two charged pions 
at a time close to its lifetime $\tau_S$ in measurements of the $D\to f K_S^0$ processes. Hence, not only $K_S^0$, but also $K_L^0$ serve as the intermediate states in the $D\to f K(t)(\to \pi^+\pi^-)$ chain decays
through the $K_S^0$-$K_L^0$ oscillation, and to their $CP$ asymmetries 
\cite{Grossman:2011zk}. The $K_S^0$ and $K_L^0$ states are linear 
combinations of the flavor eigenstates
\begin{equation}\label{eq:KSKL}
|K_{S,L}^0\rangle  =   p|K^0\rangle\mp q|\overline{K}^0\rangle,
\end{equation}
where $q/p=(1-\epsilon)/(1+\epsilon)$, and $\epsilon$ is a small
complex parameter characterizing the indirect $CP$ violation in
the kaon mixing with the magnitude
$|\epsilon|=(2.228\pm0.011)\times10^{-3}$ and the phase
$\phi_{\epsilon}=43.52\pm0.05^{\circ}$ \cite{PDG}. Let $m_{S,L}$,
$\Gamma_{S,L}$, and $\tau_{S,L}$ denote the masses,
widths, and lifetimes of $|K_{S,L}^0\rangle$, respectively. The average of
widths is then given by $\Gamma=(\Gamma_S+\Gamma_L)/2$, and the differences of widths and masses are $\Delta\Gamma\equiv\Gamma_S-\Gamma_L$ and $\Delta m\equiv m_L-m_S$, respectively.
We write the ratio between the DCS and CF amplitudes as
\begin{equation}\label{r}
 \mathcal{A}(D\rightarrow fK^0)/\mathcal{A}(D\rightarrow
  f\overline{K}^0) = r_f\,e^{i(\phi+\delta_f)},
\end{equation}
with the magnitude $r_f\propto|V_{cd}^{*}V_{us}/V_{cs}^{*}V_{ud}|
\sim\mathcal{O}(10^{-2})$, the relative strong phase $\delta_f$
that depends on final states, and the weak phase
$\phi\equiv Arg\left[-V_{cd}^{*}V_{us}/V_{cs}^{*}V_{ud} \right]
=(-6.2\pm 0.4)\times 10^{-4}$ in the SM.

\begin{figure}[!]
\includegraphics[scale=0.18]{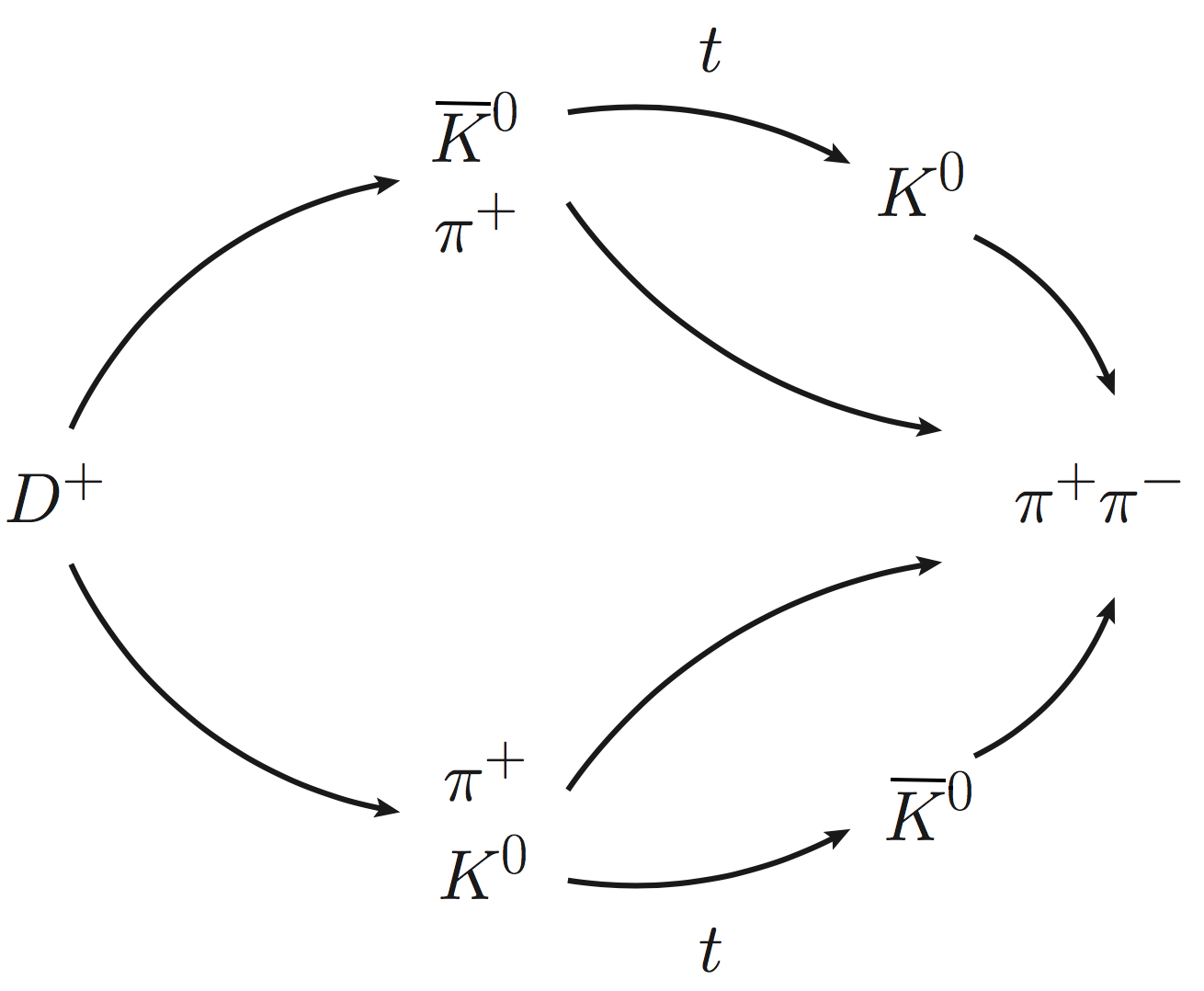}
\caption{Schematic description of the chain decay $D^+\to \pi^+K(t)(\to \pi^+\pi^-)$.
} \label{fig:amp}
\end{figure}

We consider the time-dependent $CP$ asymmetry
\begin{equation}\label{m1}
A_{CP}(t) \equiv\frac{\Gamma_{\pi\pi}(t)-\overline
\Gamma_{\pi\pi}(t)}{\Gamma_{\pi\pi}(t)+\overline\Gamma_{\pi\pi}(t)},
\end{equation}
where
\begin{equation}
\begin{split}
  \Gamma_{\pi\pi}(t)&\equiv\Gamma(D\to fK(t)(\to \pi^{+}\pi^{-})),  \\
\overline\Gamma_{\pi\pi}(t)&\equiv\Gamma(\overline D\to \overline fK(t)
(\to \pi^{+}\pi^{-})).
\end{split}
\end{equation}
Neglecting the tiny direct $CP$ asymmetry in the $K\to\pi\pi$ decays, 
namely, assuming the equality of the amplitudes 
$\mathcal{A}(\overline K^0 \to \pi^+\pi^-)=-\mathcal{A}(K^0 \to \pi^+\pi^-)$,
we derive from Eq.~(\ref{m1}), 
%
\begin{equation}\label{eq:ACPt}
 A_{CP}(t)\simeq
\left[A_{CP}^{\overline K^0}(t)+A_{CP}^{\rm dir}(t)+A_{CP}^{\rm int}(t)\right]/{D(t)},
\end{equation}
with the denominator 
$D(t)= e^{-\Gamma_St}(1-2r_f\cos\delta_f\cos\phi)+e^{-\Gamma_{L}t}|\epsilon|^{2}$.
The first term corresponds to the known $CP$ violation in the kaon mixing \cite{Grossman:2011zk},
\begin{align}\label{eq:AcpK0}
\begin{split}
A_{CP}^{\overline K^0}(t)
&=2e^{-\Gamma_St}  \mathcal{R}e(\epsilon)-2e^{-\Gamma t}
\Big[\mathcal{R}e(\epsilon)\cos(\Delta mt)
\\
&~~~~~~~+\mathcal{I}m(\epsilon)\sin(\Delta mt)\Big],
\end{split}
\end{align}
which is independent of $r_f$, i.e., of the DCS amplitude.
The second term is the direct $CP$ asymmetry originating from the
interference between the CF and DCS amplitudes,
\begin{align}
A_{CP}^{\rm dir}(t)&=e^{-\Gamma_St}\,2r_f\sin\delta_f\sin\phi.
\end{align}

The third term in Eq.~(\ref{eq:ACPt}) represents the new $CP$-violation 
effect, 
\begin{align}\label{eq:Acpint}
\begin{split}
&A_{CP}^{\rm int}(t)
=-4r_f\cos\phi\sin\delta_f\Big[e^{-\Gamma_St}\mathcal{I}m(\epsilon)
\\&~~-e^{-\Gamma t}
\Big(\mathcal{I}m(\epsilon)\cos(\Delta mt)-\mathcal{R}e(\epsilon)\sin(\Delta mt)\Big)\Big],
\end{split}
\end{align}
which is induced by the interference between the CF and DCS amplitudes of
the decays $D\to f\overline K^0(t)(\to \pi^+\pi^-)$ and
$D\to f K^0(t) (\to\pi^+\pi^-)$ with the kaon mixing.
The mechanism responsible for Eq.~(\ref{eq:Acpint})
is more complicated than for the ordinary mixing-induced $CP$
asymmetry in, for example, the $B^{0}(t)\to\pi^+\pi^-$ mode: both
the oscillation and decay take place in the mother particle in the
latter, while $A_{CP}^{\rm int}$ arises from the mother decay and the 
daughter mixing as depicted in Fig.~\ref{fig:amp}. 
$A_{CP}^{\rm int}$ 
is not a direct $CP$ asymmetry in charm decays, since it does 
not vanish as $\phi=0$. 

Compared to the SCS case, both the CF and DCS amplitudes,
being of the tree level, can be extracted from data of branching fractions
\cite{FAT,FAT2,Muller:2015lua,Biswas:2015aaa}.
A global fit to the newest data in the factorization-assisted
topological-amplitude (FAT) approach \cite{FAT} gives the parameters 
$r_{\pi^+,K^+}$ and $\delta_{\pi^+,K^+}$ for the $D^{+}\to \pi^{+}K_{S}^{0}$
and $D_{s}^{+}\to K^{+}K_{S}^{0}$ decays \cite{Wang:2017ksn}
\begin{equation}\label{eq:rdelta}
  \begin{split}
   r_{\pi^+}=-0.073\pm0.004, \qquad & \delta_{\pi^+}=-1.39\pm0.05, \\
    r_{K^+}=-0.055\pm0.002, \qquad & \delta_{K^+}=+1.45\pm0.05.
  \end{split}
\end{equation}
The solution with opposite signs of $\delta_{\pi^+,K^+}$ contributes
equivalently to branching fractions, which depend only on the cosine
of strong phases. The one presented above is preferred
by the central value of the $CP$-asymmetry data in Eq.~(\ref{eq:expDp})
in the global fit, to which the sign of strong phases is relevant.

\begin{figure}[!]
\includegraphics[scale=0.28]{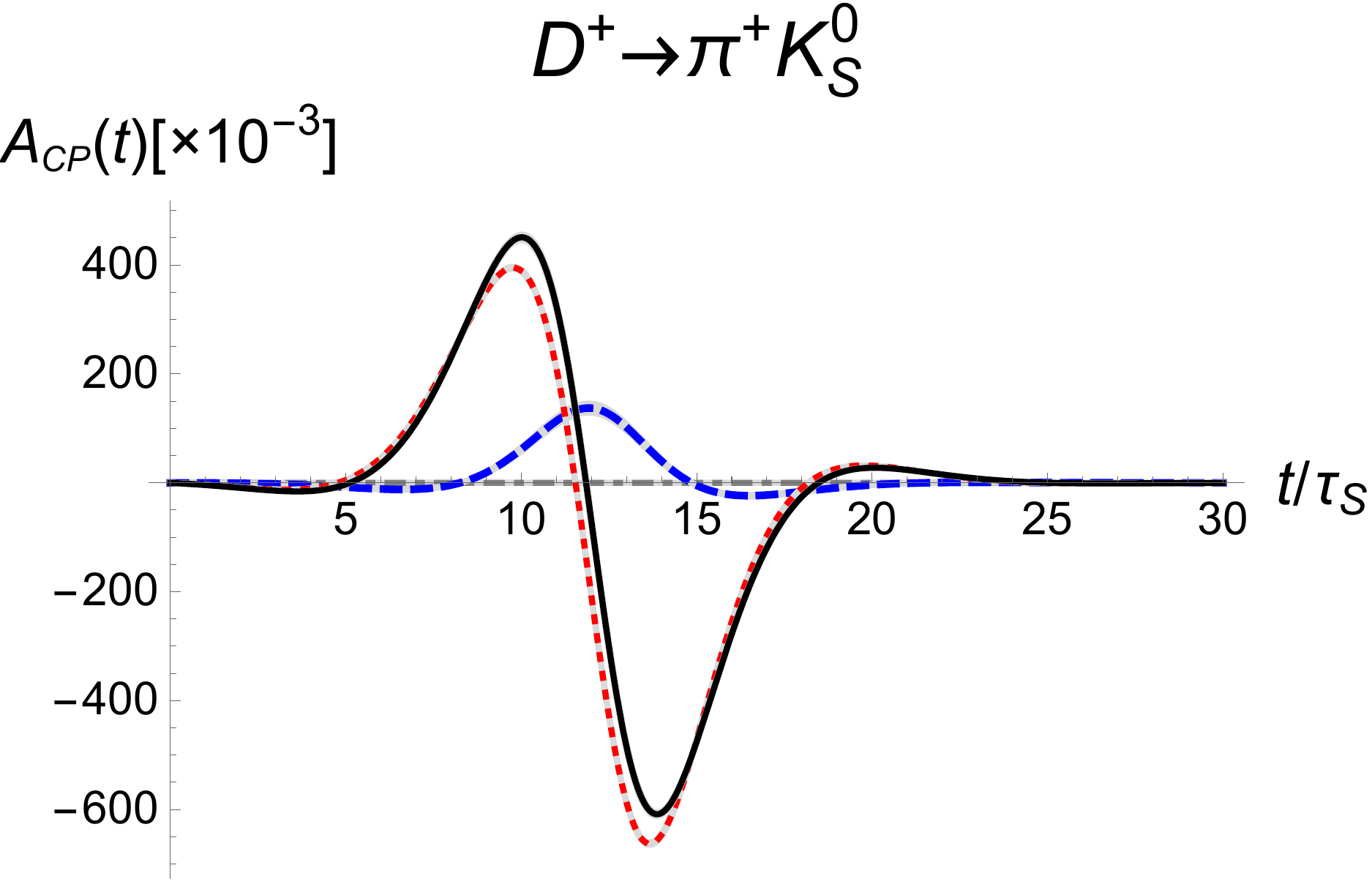}
\includegraphics[scale=0.28]{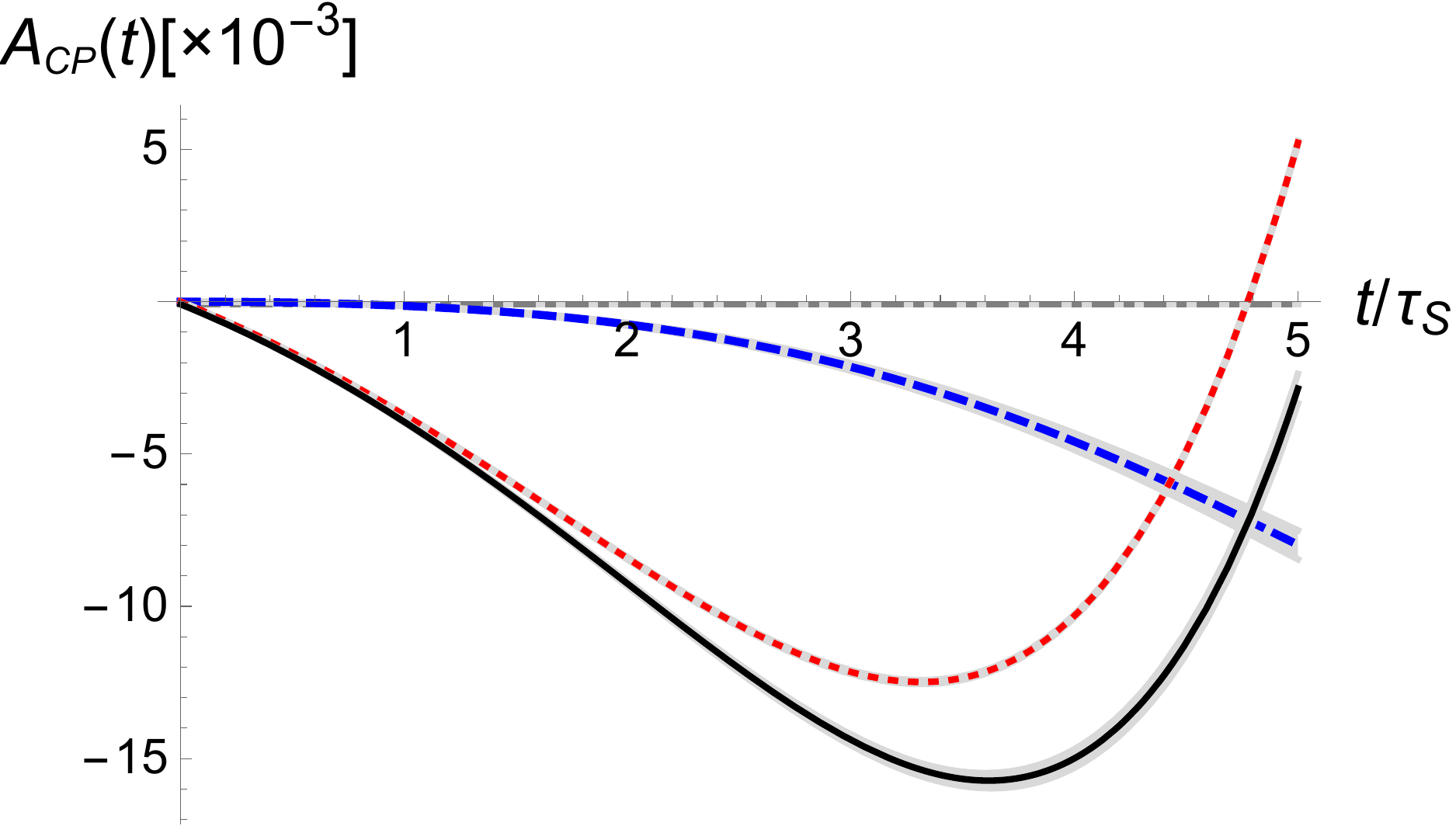}
\includegraphics[scale=0.12]{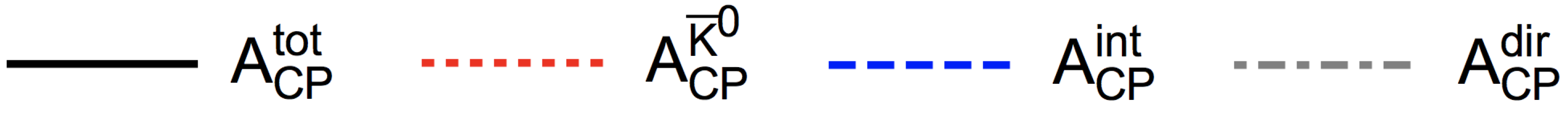}
\caption{Time-dependent $CP$ asymmetries in the $D^+\to \pi^+ K(t)(\to \pi^+\pi^-)$
decay as a function of $t/\tau_S$, with the zoomed-in plot for the small $t$ region
in the lower plot. The gray bands represent the theoretical uncertainties.
} \label{fig:ACPt}
\end{figure}

The time-dependent $CP$ asymmetries in the $D^+\to \pi^+ K(t)(\to \pi^+\pi^-)$ decay as a function of $t/\tau_S$ are displayed in Fig.~\ref{fig:ACPt}. It is found that the total $CP$
asymmetry is dominated by $A_{CP}^{\overline K^0}$, and the new effect
$A_{CP}^{\rm int}$, reaching an order of $10^{-3}$ or even $10^{-2}$ in the range $2\tau_S\lesssim t \lesssim 5\tau_S$, are experimentally accessible. 
The direct $CP$ asymmetry is too small to be seen in the plots. Hence, 
deviation of the total $CP$ asymmetry in $D\to fK_{S}^{0}$ decays from 
$A_{CP}^{\overline K^0}$ should be attributed to $A_{CP}^{\rm int}$, instead of to the direct $CP$ asymmetry.
Figure~\ref{fig:ACPt} also indicates that the total $CP$ asymmetry approaches to zero at $t=0$, because both $A_{CP}^{\overline K^0}$ and $A_{CP}^{\rm int}$
diminish at $t=0$, and $A_{CP}^{\rm dir}$ is tiny.
With the inputs in Eq.~\eqref{eq:rdelta}, the direct $CP$ asymmetries
are predicted to be
\begin{equation}\label{zz2}
\begin{split}
A_{CP}^{\rm dir}(D^+\to \pi^+K_{S}^{0}) &= (-8.6\pm0.4)\times 10^{-5}, \\
A_{CP}^{\rm dir}(D^+_s\to K^+K_{S}^{0}) &= (6.6\pm0.3)\times 10^{-5}.
\end{split}
\end{equation}
Both the forthcoming experiments, Belle II and LHCb upgrade, cannot
attain such a precision at an order of $10^{-5}$.
However, a large weak phase difference between the CF and DCS amplitudes could exist in new physics models 
\cite{Bigi:1994aw,Xing:1995jg,Lipkin:1999qz,DAmbrosio:2001mpr,Kagan:2009gb},  resulting in a larger $A_{CP}^{\rm dir}$. Therefore, an observation with 
nonvanishing $A_{CP}(t=0)$ at the Belle II and LHCb upgrade would be a
signature of new physics.

Searching for new physics through the direct $CP$ asymmetries in 
Eq.~(\ref{zz2}) might be more promising than through those in the SCS processes. 
For the latter, it is difficult to predict the $CP$ asymmetries precisely 
due to the ambiguity in estimating the penguin contributions: the 
QCD-inspired approaches do not work at the charm scale, 
and the penguin topologies cannot be extracted from data of branching 
fractions. This is the reason why predictions for $\Delta a_{CP}^{\rm dir}$ 
in the SM vary from $\mathcal{O}(10^{-4})$ to $\mathcal{O}(1\%)$
\cite{Grossman:2006jg,Buccella:1994nf,
Bigi:2011re,Artuso:2008vf,FAT,Cheng:2012wr,Brod:2011re,Pirtskhalava:2011va,
Brod:2012ud,Hiller:2012xm,Franco:2012ck,Feldmann:2012js,Khodjamirian:2017zdu,Muller:2015rna}, and cannot 
be used to discriminate new physics. 

The denominator $D(t)$ in Eq.~(\ref{eq:ACPt}) can be related to the 
$K_S^0$-$K_L^0$ asymmetry,
\begin{align}
R&\equiv {\Gamma(D\to fK_S^0)-\Gamma(D\to fK_L^0) \over \Gamma(D\to fK_S^0) +\Gamma(D\to fK_L^0)}
\nonumber\\
&=-2r_f\cos(\phi+\delta_{f})\approx-2r_{f}\cos\phi\cos\delta_{f},
\end{align}
in the limit $\phi\to 0$. The $K_S^0$-$K_L^0$ asymmetry in 
$D^{+}\to \pi^{+}K_{S,L}^{0}$ has been measured by the CLEO Collaboration with a value
$0.022\pm0.024$ \cite{He:2007aj}. The FAT approach leads to
$R(D^+\to \pi^+K_{S,L}^0)=0.025\pm0.008$, consistent with the data,
and $R(D_s^+\to K^+K_{S,L}^0)=0.012\pm0.006$ \cite{Wang:2017ksn}.
The above small results, in agreement with those derived
in the literature \cite{Cheng:2010ry,Muller:2015lua,Bhattacharya:2009ps,Gao:2014ena}, 
are due to $\delta_{\pi^+,K^+}\sim \pm\pi/2$ in Eq.~\eqref{eq:rdelta}.
That is, the term $2r_f\cos\phi\cos\delta_f$ causes an effect
at least one order of magnitude lower than $A_{CP}^{\rm int}$.

\begin{figure}[!]
    \centering
\includegraphics[scale=0.28]{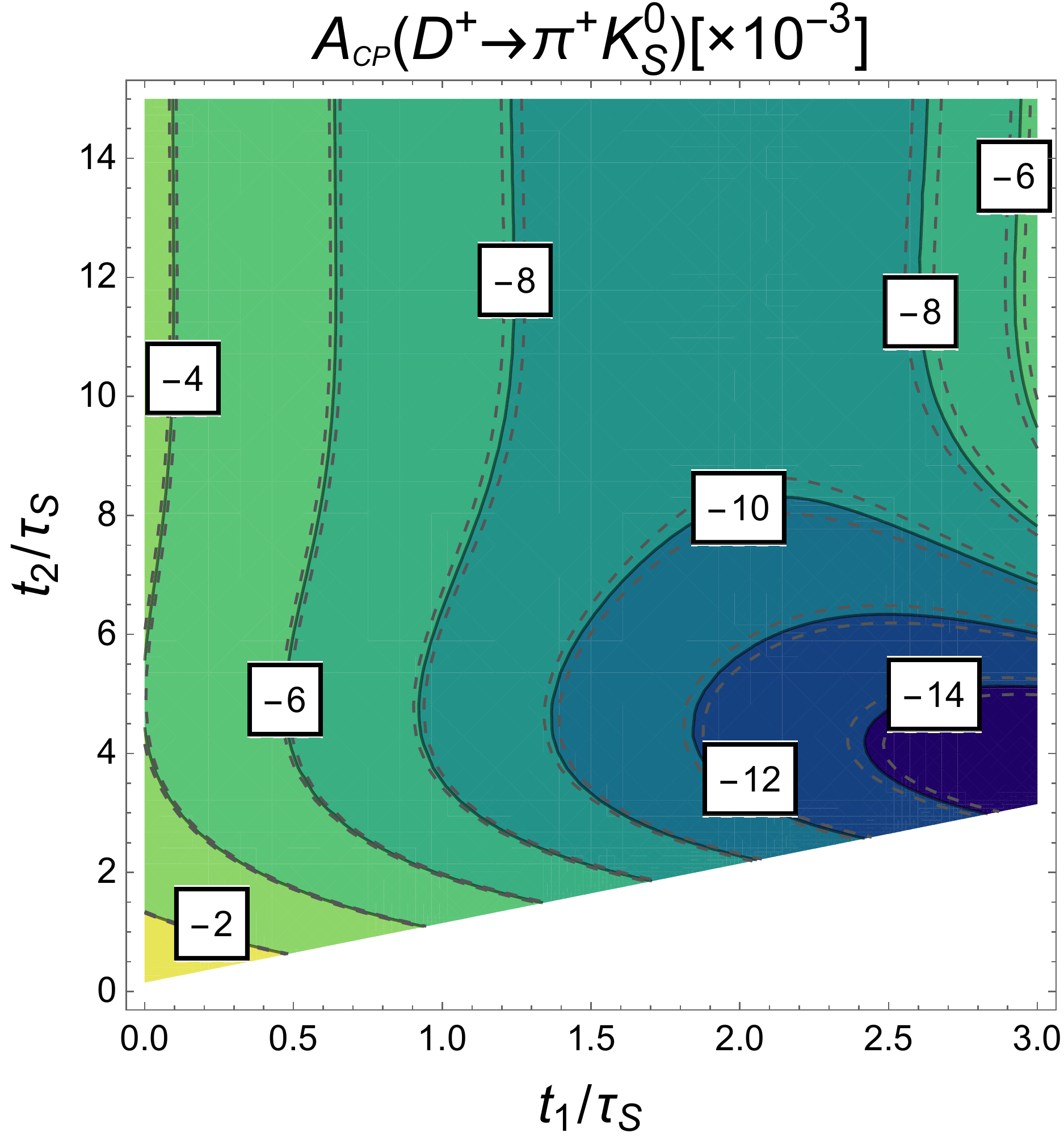}
\includegraphics[scale=0.28]{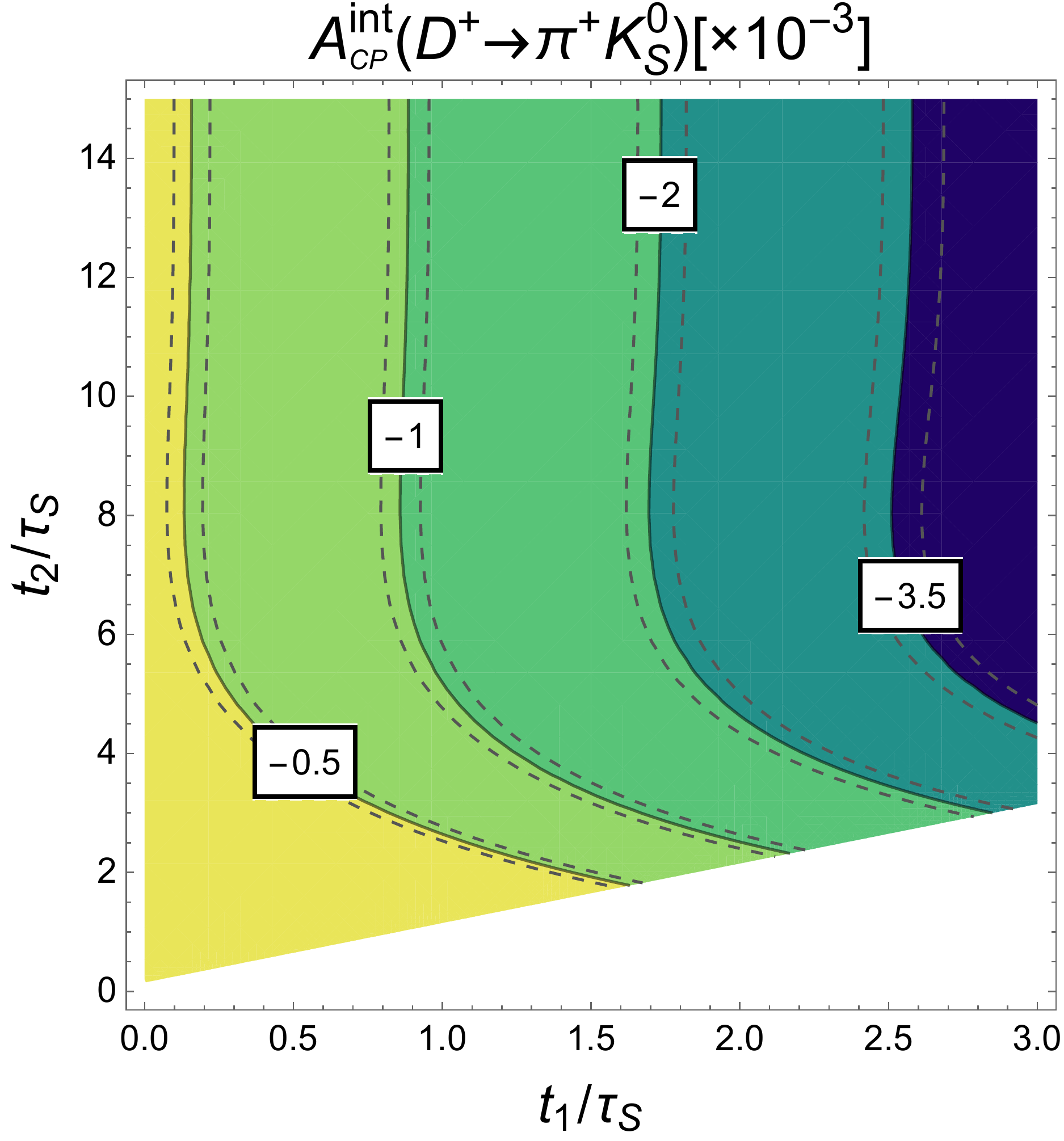}
\caption{Time-integrated $CP$ asymmetries as a function of $t_1$ and $t_2$
($t_1<t_2$) in the $D^+\to \pi^+K_S^0$ decay with the upper plot for the
total $CP$ asymmetry and the lower one for the new $CP$-violation effect.
The dashed lines indicate the theoretical uncertainties of our predictions.
} \label{fig:ACPt1t2}
\end{figure}

Measurements of $CP$ 
asymmetries depend on time intervals selected in
experiments. To obtain a time-integrated $CP$ asymmetry defined by
\begin{align}\label{time}
A_{CP}=\frac{\int_0^{\infty}F(t)[\Gamma_{\pi\pi}(t)-\overline
\Gamma_{\pi\pi}(t)]dt}{\int_0^{\infty}F(t)
[\Gamma_{\pi\pi}(t)+\overline \Gamma_{\pi\pi}(t)]dt},
\end{align}
a function of time, $F(t)$, is introduced to take into account
relevant experimental effects, such as detecting efficiencies and kaon
energies. We adopt the approximation with $F(t)=1$ in the interval
$[t_1,t_2]$ and $F(t)=0$ elsewhere \cite{Grossman:2011zk}. 
Equation~(\ref{time}) then yields
\begin{eqnarray}\label{eq:Acpt1t2}
& A_{CP}(t_1,t_2)= \frac{\int^{t_2}_{t_1}
 \Big[A_{CP}^{\overline K^0}(t)+A_{CP}^{\rm dir}(t)+A_{CP}^{\rm int}(t)\Big]dt}
 {\int^{t_2}_{t_1}D(t)dt}\nonumber\\
& ~~~~~~~~~~~~~~~=\frac{2\mathcal{R}e(\epsilon)-4\mathcal{I}m(\epsilon)
r_f\cos\phi\sin\delta_f}{1-2r_f\cos\delta_f\cos\phi}\Bigg[1-
  \nonumber\\
  &\frac{\big[c(t_1)-c(t_2)\big]
            + \frac{\mathcal{I}m(\epsilon)+2\mathcal{R}e(\epsilon)r_f\cos\phi\sin\delta_f}
                      {\mathcal{R}e(\epsilon)-2\mathcal{I}m(\epsilon)r_f\cos\phi\sin\delta_f}
               \big[s(t_1)-s(t_2)\big]}
  { \tau_S\Gamma (1+x^2)(e^{-\Gamma_St_1}-e^{-\Gamma_St_2})}
\Bigg]
\nonumber\\
&~~~+2r_f\sin\delta_f\sin\phi,
\end{eqnarray}
where $x\equiv\Delta m/\Gamma$, $c(t)=e^{-t \Gamma}[\cos(\Delta m t)-x\, \sin(\Delta m t)]$, and $s(t)=e^{-t \Gamma}[x \cos(\Delta m t)+ \sin(\Delta m t)]$. In the second and
third lines the terms proportional to $r_{f}$ stand for the new effect
$A_{CP}^{\rm int}$, and those without $r_{f}$ for the $CP$
violation in the neutral kaon system.
The last term, being independent of $t_{1,2}$, corresponds to the direct $CP$
asymmetry. The time-integrated $CP$ asymmetries in the $D^+\to \pi^+K_S^0$ 
decays are exhibited in Fig.~\ref{fig:ACPt1t2} with the upper plot for the total $CP$
asymmetry and the lower one for the new effect. Both quantities are relatively
large in some ranges of $t_1$ and $t_2$, suggesting the favorable time
intervals for experimental investigations of these $CP$ asymmetries.

With the same approximation as in \cite{Grossman:2009mn}
for the limit of $t_1\ll \tau_S \ll t_2 \ll \tau_L$, we get
\begin{align}\label{eq:t1t2limit}
 &~~~A_{CP}(t_1\ll \tau_S\ll t_2 \ll \tau_L)
 \\&
\simeq  \frac{-2\mathcal{R}e(\epsilon)
+2r_f\sin\phi\sin\delta_f-4\mathcal{I}m(\epsilon)r_f\cos\phi\sin\delta_f}
 {1-2r_f\cos\phi\cos\delta_f}.\nonumber
\end{align}
In the absence of the DCS contributions, i.e., $r_f=0$, the above formula reduces 
to $-2\mathcal{R}e(\epsilon)$ derived in 
\cite{Lipkin:1999qz,Grossman:2011zk,Bigi:1994aw,Bianco:2003vb,Xing:1995jg,Ko:2012pe}.
The effect of $A_{CP}^{\rm int}$, namely, the third term in the numerator
of Eq.~(\ref{eq:t1t2limit}) was missed in the study of the $CP$ asymmetry in
$D^+\to \pi^+K^0_S$ by the Belle Collaboration \cite{Ko:2012pe}.
The direct $CP$ violation has to be extracted by
subtracting the kaon-mixing and new effects from the total $CP$ asymmetry.  
The sum of the latter two effects in Eq.~(\ref{eq:t1t2limit}) is predicted to be 
$(-3.57\pm0.05)\times10^{-3}$. 
The direct $CP$ violation $(-0.06\pm1.15)\times10^{-3}$
is then obtained from the Belle data in Eq.~(\ref{eq:expDp}),
consistent with our prediction in Eq.~(\ref{zz2}).
The $D^+\to \pi^+K_S^0$ and $D^+_s\to K^+K_S^0$ decays have been employed to
cancel the systematic asymmetries from the production and detection at the LHCb
for the measurements of $CP$ violation in the SCS processes
\cite{Aaij:2013ula,Aaij:2014gsa,Aaij:2014qec,Aaij:2016dfb}. The working
assumption is that there is no sizable $CP$ violation other than the one from
the kaon mixing in the $D^+\to \pi^+K_S^0$ and $D^+_s\to K^+K_S^0$ decays.
However, $A_{CP}^{\rm int}$ observed here is of the same order as the direct
$CP$ asymmetries in the SCS processes, which are expected to be
$\mathcal{O}(10^{-3})$ or $\mathcal{O}(10^{-4})$. That is, the effect of
$A_{CP}^{\rm int}$ has to be considered in these measurements as well.

\begin{figure}[t!]
\centering
\includegraphics[scale=0.3]{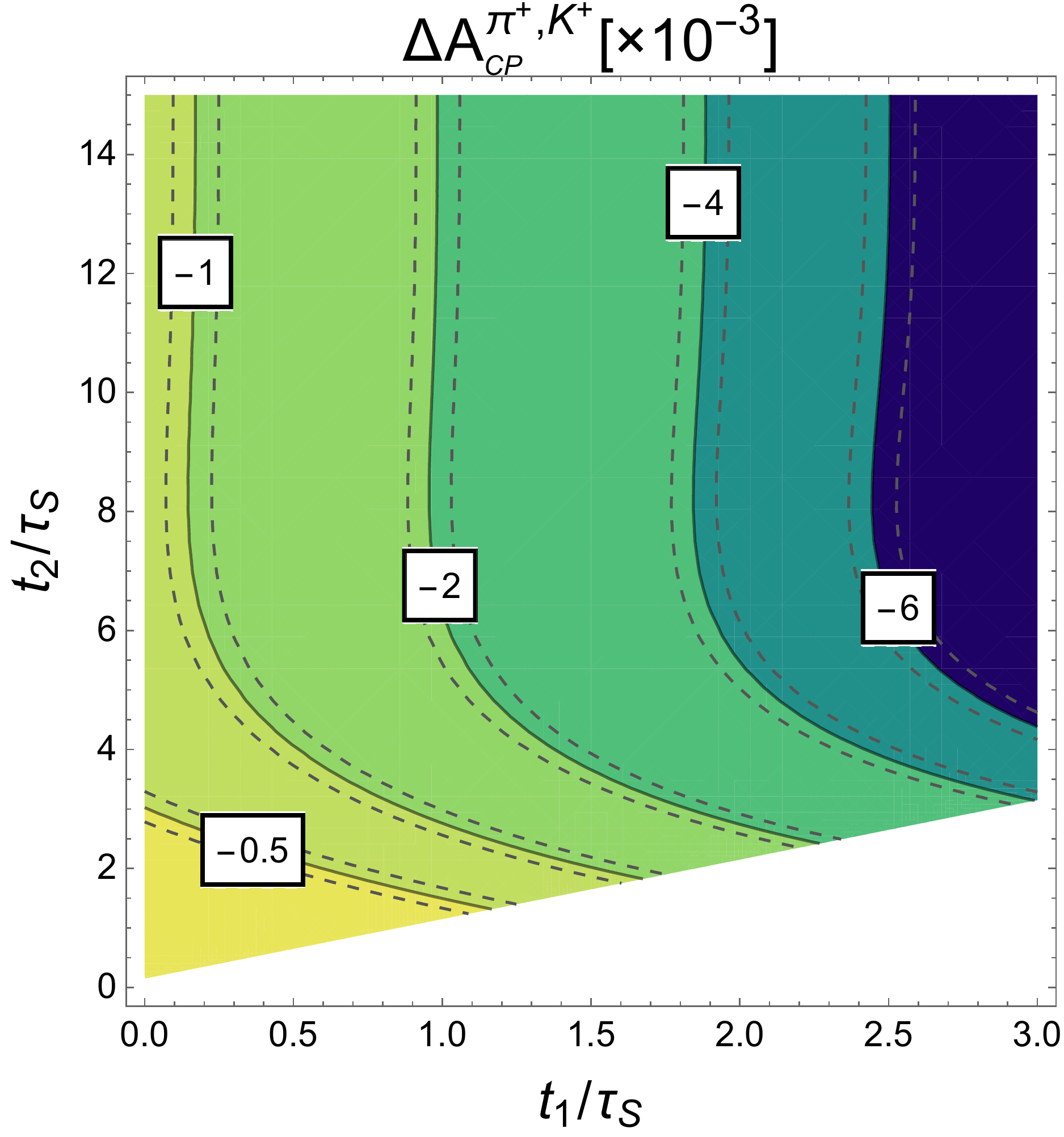}
\caption{Same as FIG.~\ref{fig:ACPt1t2} but for $\Delta A_{CP}^{\pi^+,K^+}$.}\label{fig:DeltaAcp}
\end{figure}

To verify the new $CP$-violation effect, we propose an
observable, the difference of the time-integrated $CP$ asymmetries in the
$D^+\to \pi^+K_S^0$ and $D_s^+\to K^+K_S^0$ modes,
\begin{align}\label{eq:DeltaAcp}
&\Delta  A_{CP}^{\pi^+,K^+}\equiv
A_{CP}^{D^+\to \pi^+K_S^0}(t_1,t_2)-A_{CP}^{D^+_s\to K^+K_S^0}(t_1,t_2)
\nonumber\\
&\simeq A_{CP}^{\text{int},D^+\to \pi^+K_S^0}(t_{1},t_{2})
-A_{CP}^{\text{int},D^+_s\to K^+K_S^0}(t_1,t_2).
\end{align}
Our global-fit analysis indicates that the new effect is the most significant 
in this observable. The $CP$ violation in the kaon mixing, being mode-independent as implied by Eq.~(\ref{eq:AcpK0}), is canceled in the above difference, and the direct $CP$ violation is negligible.
The new effect survives in $\Delta A_{CP}^{\pi^+,K^+}$
according to the following model-independent argument.
The topological diagrams of the CF and DCS amplitudes in these two decays are exchanged to each other under the flavor $SU(3)$ symmetry,
$\mathcal{A}(D^{+}\to \pi^{+}K^{0})/V_{us}V_{cd}^{*}=
\mathcal{A}(D_{s}^{+}\to K^{+}\overline K^{0})/V_{ud}V_{cs}^{*}=C+A$
and $\mathcal{A}(D^{+}\to \pi^{+}\overline K^{0})/V_{ud}V_{cs}^{*}
=\mathcal{A}(D_{s}^{+}\to K^{+} K^{0})/V_{us}V_{cd}^{*}=T+C$,
with the color-favored tree-emission diagram $T$, the
color-suppressed tree-emission diagram $C$, and the $W$-annihilation
diagram $A$ \cite{Chau:1982da,Chau:1986du,Bhattacharya:2008ss}. The relative strong phases $\delta_f$ in these two modes are, thus, opposite in sign, as shown in Eq.~(\ref{eq:rdelta}), so that the new effects are constructive in $\Delta A_{CP}^{\pi^+,K^+}$. The dependencies of $\Delta A_{CP}^{\pi^+,K^+}$ on $t_1$ and $t_2$ are displayed in Fig.~\ref{fig:DeltaAcp}. It is seen that this observable is of the order of $10^{-3}$ in most of the time intervals, and increases with $t_1$.

The effect of $A_{CP}^{\rm int}$ are measurable
in the forthcoming experiments. The precision of Belle II measurements
on the $CP$ asymmetry in $D^{+}\to \pi^{+}K_{S}^{0}$ can attain
$3\times10^{-4}$ at 50 ab$^{-1}$ \cite{Schwartz:2017gni}. In the
LHCb upgrade, the error bar of $\Delta A_{CP}$ defined in Eq.~(\ref{data})
would be reduced to $1.2\times10^{-4}$ at 50 fb$^{-1}$ \cite{Bediaga:2012py}. The signal yields of $D^{+}\to \pi^{+}K_{S}^{0}$ and
$D_{s}^{+}\to K^{+}K_{S}^{0}$ are of the same order as of
$D^{0}\to K^{+}K^{-}$ and $\pi^{+}\pi^{-}$ \cite{Aaij:2014qec,Aaij:2016cfh}.
It is then expected that the precision of $\Delta A_{CP}^{\pi^{+},K^{+}}$ can also reach $\mathcal{O}(10^{-4})$ at the LHCb upgrade, and that $\Delta A_{CP}^{\pi^{+},K^{+}}
\sim\mathcal{O}(10^{-3})$ is accessible at both the Belle II and LHCb upgrade.

In this Letter, we have studied the time-dependent and time-integrated
$CP$ asymmetries in the $D\to f K_{S}^{0}(\to \pi^{+}\pi^{-})$ chain
decays. A new $CP$-violation effect was identified in
these processes, which is induced by the interference between the
CF and the DCS amplitudes with the $K^{0}$-$\overline K^{0}$ mixing.
Compared to the SCS processes, both the CF and DCS amplitudes, occurring
at the tree level, can be extracted from the data of branching fractions.
Therefore, their $CP$ asymmetries can be estimated more accurately,
and have been shown to be as large as $10^{-3}$ in the
$D^+\to \pi^+K_S^0$ and $D_s^+\to K^+K_S^0$ modes.
Nevertheless, its effect has been missed in the literature.
To reveal this new $CP$-violation effect, we have proposed an observable,
the difference of the $CP$ asymmetries in the $D^+\to \pi^+K_S^0$ and
$D_s^+\to K^+K_S^0$ decays accessible at Belle II
and LHCb. In addition, the direct $CP$ asymmetries used to search for new physics can be determined either by subtracting the kaon-mixing and DCS interference effects from total $CP$ asymmetries, or by the time-dependent measurements of $CP$ violation in these processes.

We are grateful to Peng-Fei Guo for numerical checks, and to Ji-bo He, Ying Li, Cai-Dian L\"u, Xiao-Rui Lyu, Cheng-Ping Shen, Liang Sun and Wei Wang for useful discussions.
This work was supported in part by the National Natural Science
Foundation of China under Grants No. 11347027, 11505083 and U1732101, and by the Ministry of Science and Technology of R.O.C. under Grant No. MOST-104-2112-M-001-037-MY3.


\end{document}